\documentclass{llncs}
\usepackage{graphics}

\newcommand{\comment}[1]{}

\def \lket {|}
\def \rket {\rangle}

\newcommand{\ket}[1]{\lket #1\rket}

\def\cents{\hbox{\rm\rlap/c}}
\def\bbbc{{\mathchoice {\setbox0=\hbox{$\displaystyle\rm C$}\hbox{\hbox
to0pt{\kern0.4\wd0\vrule height0.9\ht0\hss}\box0}}
{\setbox0=\hbox{$\textstyle\rm C$}\hbox{\hbox
to0pt{\kern0.4\wd0\vrule height0.9\ht0\hss}\box0}}
{\setbox0=\hbox{$\scriptstyle\rm C$}\hbox{\hbox
to0pt{\kern0.4\wd0\vrule height0.9\ht0\hss}\box0}}
{\setbox0=\hbox{$\scriptscriptstyle\rm C$}\hbox{\hbox
to0pt{\kern0.4\wd0\vrule height0.9\ht0\hss}\box0}}}}

\newtheorem{Hypothesis}{Hypothesis}

\begin{document}
\title{Improved constructions of quantum automata}
\author{Andris Ambainis and Nikolajs Nahimovs\thanks{Supported by 
University of Latvia research project Y2-ZP01-100.}
}
\institute{Department of Computer Science, University of Latvia,
Raina bulv. 19, Riga, LV-1586, Latvia,
{\tt andris.ambainis@lu.lv, kolja.nahimov@gmail.com}.} 
\date{}

\maketitle
\abstract{We present a simple construction of quantum automata
which achieve an exponential advantage over classical finite automata.
Our automata use $\frac{4}{\epsilon} \log 2p  + O(1)$ states to recognize a language
that requires $p$ states classically. The construction is
both substantially simpler and achieves a better constant in the
front of $\log p$ than the previously known construction of \cite{AF}.

Similarly to \cite{AF}, our construction is by a probabilistic
argument. We consider the possibility to derandomize it and
present some results in this direction.
}

\section{Introduction}

Quantum finite automata are a mathematical model for quantum computers
with limited memory. A quantum finite automaton has a finite state
space and applies a sequence of transformations, corresponding
to the letter of the input word to this state space. At the end,
the state of the quantum automaton is measured and the input word
is accepted or rejected, depending on the outcome of the measurement.

Most commonly, finite automata (including quantum finite automata)
are studied in 1-way model where the transformations corresponding
to the letters of the input word are applied in the order of the letters
in the word, from the left to the right. (More general 2-way models \cite{KW}
allow the order of the transformations to depend on the results 
of the previous transformations.) 

For 1-way model (which we consider 
the most natural model in the quantum setting),
the set of languages (computational problems)
that can be recognized (computed) by a quantum automaton
is the same for classical automata\footnote{
More precisely,
this is true for sufficiently general models of quantum
automata, such as one proposed in \cite{BMP} or \cite{Ciamarra}.
There are several results 
claiming that quantum automata are weaker than classical 
(e.g. \cite{KW,AKV,ANTV}) but this is an artifact of restrictive models
of quantum automata being used.}. 
However, quantum automata
can be exponentially more space-efficient than classical 
automata \cite{AF}. This is one of only two results that show
an exponential advantage for quantum algorithms in space complexity.
(The other is the recent exponential separation for online
algorithms by Le Gall \cite{LeGall}.)

Our first result is an improved exponential separation between quantum
and classical finite automata, for the same computational problem
as in \cite{AF}. The construction in \cite{AF} is quite inefficient.
While it produces an example where classical automata require $p$ states
and quantum automata require $C \log p$ states, the constant $C$ is 
fairly large. In this paper, we provide a new construction with 
a better constant and, also, a much simpler analysis. (A detailed
comparison between our results and \cite{AF} is given in section 
\ref{sec:overview}.)

Second, both construction of QFAs in \cite{AF} and this paper are 
probabilistic. That is, they employ a sequence of parameters that
are chosen at random and hardwired into the QFA. 
In the last section, we give two non-probabilistic constructions 
of QFAs for the same language.
The first of them gives QFAs with $O(\log p)$ states
but its correctness is only shown by numerical experiments.
The second construction gives QFAs with $O(\log^{2+\epsilon} p)$
states but is provably correct.

\section{Definitions}

\subsection{Quantum finite automata}

We consider 1-way quantum finite automata (QFA) 
as defined in \cite{CM}. Namely, a 1-way QFA is a tuple 
$M=(Q, \Sigma, \delta, q_0, Q_{acc}, Q_{rej})$
where $Q$ is a finite set of states, $\Sigma$ is
an input alphabet, $\delta$ is a transition function, 
$q_0\in Q$ is a starting state,
$Q_{acc}$ and $Q_{rej}$ are sets of accepting and
rejecting states and $Q=Q_{acc}\cup Q_{rej}$. 
$\cents$ and $\$$ are symbols that do not belong to $\Sigma$.
We use $\cents$ and $\$$ as the left and the right endmarker, respectively.
The {\em working alphabet} of $M$ is $\Gamma=\Sigma\cup\{\cents, \$\}$.

A superposition of $M$ is any element of $l_2(Q)$
(the space of mappings from $Q$ to $\bbbc$ with $l_2$ norm).
For $q\in Q$, $\ket{q}$ denotes the unit vector with
value 1 at $q$ and 0 elsewhere. 
All elements of $l_2(Q)$ can be expressed as linear
combinations of vectors $\ket{q}$.
We will use $\psi$ to denote elements of $l_2(Q)$.

The transition function $\delta$ maps $Q\times\Gamma\times Q$ to $\bbbc$.
The value $\delta(q_1, a, q_2)$ is the amplitude of $\ket{q_2}$ in
the superposition of states to which $M$ goes from $\ket{q_1}$ after
reading $a$. For $a\in\Gamma$, $V_a$ is a linear transformation on $l_2(Q)$
defined by
\begin{equation}
\label{E1}
V_a(\ket{q_1})=\sum_{q_2\in Q}\delta(q_1, a, q_2)\ket{q_2}.
\end{equation}
We require all $V_a$ to be unitary.


The computation of a QFA starts in the superposition
$\ket{q_0}$. Then transformations corresponding to 
the left endmarker $\cents$, the letters of the input word $x$ and
the right endmarker $\$$ are applied. 
The transformation corresponding to $a\in\Gamma$ is just
$V_a$. If the superposition before reading $a$ is $\psi$,
then the superposition after reading $a$ is $V_a(\psi)$.

After reading the right endmarker, the current state 
$\psi$ is observed with respect to 
the observable $E_{acc}\oplus E_{rej}$
where $E_{acc}=span\{\ket{q} : q\in Q_{acc}\}$,
$E_{rej}=span\{\ket{q} : q\in Q_{rej}\}$.
This observation gives $x\in E_{i}$ with the probability
equal to the square of the projection of $\psi$ to $E_i$.
After that, the superposition collapses to this projection.

If we get $\psi\in E_{acc}$, the input is accepted.
If $\psi\in E_{rej}$, the input is rejected.

{\bf Another definition of QFAs.}
Independently of \cite{CM}, quantum automata were 
introduced in \cite{KW}. There is one difference between
these two definitions. In \cite{KW}, a QFA is observed after
reading each letter (after doing each $V_a$).
In \cite{CM}, a QFA is observed only after all letters have been
read.
The definition of \cite{KW} is more general.
But, in this paper, we follow the definition of \cite{CM} because
it is simpler and sufficient
to describe our automaton. 

\subsection{Unitary transformations}

We use the following theorem from linear algebra.
\begin{theorem}
\label{thm:la}
Let $\alpha_1$, $\ldots$, $\alpha_m$ be such that
$|\alpha_1|^2+\ldots+|\alpha_m|^2=1$.
Then, 
\begin{enumerate}
\item
there is a unitary transformation $U_1$ such that
$U_1\ket{q_1}=\alpha_1\ket{q_1}+\ldots+\alpha_m\ket{q_m}$.
\item
there is a unitary transformation $U_2$ such that,
for all $i\in\{1, \ldots, m\}$, $U_2\ket{q_i}$ is equal to
$\alpha_i\ket{q_1}$ plus some combination of $\ket{q_2}$, $\ldots$, $\ket{q_m}$. 
\end{enumerate}
\end{theorem}

In the second case, we also have 
\[ U_2 (\alpha_1\ket{q_1}+\ldots+\alpha_m\ket{q_m})=\ket{q_1} .\]

\section{Space-efficient quantum automaton}

\subsection{Summary of results}
\label{sec:overview}

Let $p$ be a prime. We consider the language  
$L_p$ = \{ $a^i$ $|$ $i$ is divisible by $p$ \}.
It is easy to see that any deterministic 1-way finite 
automaton recognizing $L_p$ has at least $p$ states.
However, there is a much more efficient QFA!
Namely, Ambainis and Freivalds \cite{AF} have shown that $L_p$ can
be recognized by a QFA with $O(\log p)$ states.

The big-O constant in this result depends on the required probability
of correct answer. For $x\in L_p$, the answer is always correct with
probability 1. For $x\notin L_p$, \cite{AF} give
\begin{itemize}
\item
a QFA with $16\log p$ states that is correct with probability
at least 1/8 on inputs $x\notin L_p$.
\item
a QFA with $poly(\frac{1}{\epsilon}) \log p$ states that is
correct with probability at least $1-\epsilon$ on inputs $x\notin L_p$
(where $poly(x)$ is some polynomial in $x$).
\end{itemize}

In this paper, we present a simpler construction of QFAs that achieves
a better big-O constant.

\begin{theorem}
\label{T6}
For any $\epsilon>0$, there is a QFA with  $4 \frac {\log {2p}}{\epsilon}$ 
states recognizing
$L_p$ with probability at least $1-\epsilon$.
\end{theorem}

\subsection{Proof of Theorem \ref{T6}}

Let $U_k$, for $k\in\{1, \ldots, p-1\}$, be a quantum automaton 
with a set of states $Q = \{q_0, q_1\}$, 
a starting state $\ket{q_0}$,
$Q_{acc}=\{q_{0}\}$, $Q_{rej}=\{ q_{1}\}$.
The transition function is defined as follows.
Reading $a$ maps $\ket{q_0}$ to $\cos\phi\ket{q_0} + \sin\phi\ket{q_1}$
and $\ket{q_1}$ to $-\sin\phi\ket{q_0} + \cos\phi\ket{q_1}$
where $\phi=\frac{2\pi k}{p}$.
(It is easy to check that this transformation is unitary.)
Reading $\cents$ and $\$$ leaves $\ket{q_0}$ and $\ket{q_1}$ unchanged.

\begin{lemma}
\label{L3}
After reading $a^j$, the state of $U_k$ is
\[ \cos\left(\frac{2\pi j k}{p}\right)\ket{q_0}
+\sin\left(\frac{2\pi j k}{p}\right)\ket{q_1}.\]
\end{lemma}

\noindent
{\bf Proof.}
By induction. 
\qed

If $j$ is divisible by $p$, then $\frac{2\pi j k}{p}$
is a multiple of $2\pi$, $\cos(\frac{2\pi j k}{p})=1$,
$\sin(\frac{2\pi j k}{p})=0$, reading $a^j$ maps
the starting state $\ket{q_0}$ to $\ket{q_0}$.
Therefore, we get an accepting state with probability 1.
This means that all automata $U_k$ accept words in $L$ with probability 1.

Let $k_1, \ldots, k_d$ be a sequence of $d=c \log p$ numbers. 
We construct an automaton $U$ by combining $U_{k_1}$,
$\ldots$, $U_{k_d}$.
The set of states consists of $2d$ states $q_{1,0}$, $q_{1,1}$,
$q_{2,0}$, $q_{2,1}$, $\ldots$, $q_{d,0}$, $q_{d,1}$.
The starting state is $q_{1, 0}$.

The transformation for left endmarker $\cents$ is such that $V_{\cents}(\ket{q_{1,0}}) = \ket{\psi_{0}}$ where
\[ \ket{\psi_{0}}=\frac{1}{\sqrt{d}} (\ket{q_{1,0}}+\ket{q_{2,0}}+ \ldots
\ket{q_{d,0}}) .\]
This transformation exists by first part of Theorem \ref{thm:la}.
The transformation for $a$ is defined by
\[ V_a(\ket{q_{i,0}}) = \cos\frac{2k_i \pi}{p} \ket{q_{i,0}} + \sin\frac{2k_i \pi}{p} \ket{q_{i,1}} ,\]
\[ V_a(\ket{q_{i,1}}) = -\sin\frac{2k_i \pi}{p} \ket{q_{i,0}} + \cos\frac{2k_i \pi}{p} \ket{q_{i,1}} .\]
The transformation $V_{\$}$ is as follows. 
The states $\ket{q_{i, 1}}$ are left unchanged. 
On the states $\ket{q_{i,0}}$, 
$V_{\$}\ket{q_{i,0}}$ is $\frac{1}{\sqrt{d}}\ket{q_{1, 0}}$ plus some other state
(part 2 of Theorem \ref{thm:la}, applied to $\ket{q_{1, 0}}$, $\ldots$, 
$\ket{q_{d,0}}$).
In particular,
\[ V_{\$} \ket{\psi_{0}}=\ket{q_{1,0}} .\]
The set of accepting states $Q_{acc}$ consists of one state $q_{1,0}$.
All other states $q_{i,j}$ belong to $Q_{rej}$.

\begin{claim}
If the input word is $a^j$ and $j$ is divisible by $p$,
then $U$ accepts with probability 1.
\end{claim}

\proof
The left endmarker maps the starting state to $\ket{\psi_0}$.
Reading $j$ letters $a$ maps each $\ket{q_{i,0}}$ to itself
(see analysis of $U_k$).
Therefore, the state $\ket{\psi_0}$ which consists of 
various $\ket{q_{i,0}}$ is also mapped to itself.
The right endmarker maps $\ket{\psi_0}$ to $\ket{q_{1,0}}$ which
is an accepting state.
\qed

\begin{claim}
If the input word is $a^j$, $j$ not divisible by $p$,
$U$ accepts with probability
\begin{equation}
\label{eqn}
\frac{1}{d^2} \left( \cos\frac{2\pi k_1 j}{p}+\cos\frac{2\pi k_2 j}{p}+\ldots+
\cos\frac{2\pi k_d j}{p}\right)^2 .
\end{equation}
\end{claim}

\proof
By Lemma \ref{L3}, $a^j$ maps $\ket{q_{i,0}}$ to
$\cos\frac{2\pi k_i j}{p} \ket{q_{i,0}}+\sin\frac{2\pi k_i j}{p} \ket{q_{i,1}}$.
Therefore, the state before reading the right endmarker $\$$ is
\[ \frac{1}{\sqrt{d}} \sum_{i=1}^d ( \cos\frac{2\pi k_i j}{p} \ket{q_{i,0}}+\sin\frac{2\pi k_i j}{p} \ket{q_{i,1}} ) .\]
The right endmarker maps each $\ket{q_{i,0}}$ to $\frac{1}{\sqrt{d}}\ket{q_{1,0}}$ plus superposition of other basis states.
Therefore, the state after reading the right endmarker $\$$ is
\[ \frac{1}{d} \sum_{i=1}^d  \cos\frac{2\pi k_i j}{p} \ket{q_{1,0}} \]
plus other states $\ket{q_{i,j}}$. 
Since $\ket{q_{1,0}}$ is the only accepting state,
the probability of accepting is the square of the coefficient of $\ket{q_{1,0}}$.
This proves the lemma.
\qed

We use the following theorem from probability theory (variant of Azuma's theorem\cite{MR}).

\begin{theorem}
\label{thm:azuma}
Let $X_1, \ldots, X_d$ be independent random variables 
such that $E[X_i]=0$ and the value of $X_i$ is always
between -1 and 1. Then,
\[ Pr[ | \sum_{i=1}^d X_i | \geq \lambda] \leq 2 e^{-\frac{\lambda^2}{2d}} .\]
\end{theorem}

We apply this theorem as follows. Fix $j\in\{1, \ldots, p-1\}$.
Pick each of $k_1, \ldots, k_d$ randomly from $\{0, \ldots, p-1\}$. 
Define $X_i=\cos \frac{2\pi k_i j}{p}$.
We claim that $X_i$ satisfy the conditions of theorem.
Obviously, the value of $\cos$ function is between -1 and 1.
The expectation of $X_i$ is 
\[ E[X_i]=\frac{1}{p} \sum_{k=0}^{p-1} \cos \frac{2\pi k j}{p} \]
since $k_i=k$ for each $k\in\{0, \ldots, p-1\}$ with probability $1/p$.
We have $\cos \frac{2\pi k j}{p}=\cos \frac{2\pi (k j \bmod p)}{p}$ 
because $\cos (2\pi+x)=\cos x$. 
Consider the numbers 0, $j$, $2j \bmod p$, $\ldots$, $(p-1)j\bmod p$.
They are all distinct.
(Since $p$ is prime, $kj=k'j (\bmod p)$ implies $k=k'$.)
Therefore, the numbers 0, $j$, $2j \bmod p$, $\ldots$, $(p-1)j\bmod p$
are just $0, 1, \ldots, p-1$ in a different order.
This means that the expectation of $X_i$ is 
\[ E[X_i]=\frac{1}{p} \sum_{k=0}^{p-1} \cos \frac{2\pi k}{p} .\]
This is equal to 0.

By equation (\ref{eqn}), the probability of accepting $a^j$ is
$\frac{1}{d^2} (X_1+\ldots+X_d)^2$.
To achieve 
\[ \frac{1}{d^2} (X_1+\ldots+X_d)^2 \leq \epsilon ,\]
we need $|X_1+\ldots+X_d| \leq \sqrt{\epsilon} d$.
By Theorem \ref{thm:azuma}, the probability that this does not happen 
is at most $2 e^{-\frac{\epsilon d}{2}}$.

There are $p-1$ possible inputs not in $L$:
$a^1$, $\ldots$, $a^{p-1}$.
The probability that one of them gets accepted with
probability more than $\epsilon$ is
at most $2(p-1) e^{-\frac{\epsilon d}{2}}$.
If
\begin{equation}
\label{eq:2} 
2(p-1) e^{-\frac{\epsilon d}{2}} < 1 ,
\end{equation}
then there is at least one choice of $k_1, \ldots, k_d$ 
for which $U$ does not accept any of $a^1$, $\ldots$, $a^{p-1}$ 
with probability more than $\epsilon$.
The equation (\ref{eq:2}) is true if we take 
$d =2\frac{\log 2p}{\epsilon}$.
The number of states for $U$ is 
$4\frac{\log 2p}{\epsilon}$.
\qed

\section{Explicit constructions of QFAs}

In the previous section, we proved what for every $\epsilon>0$ and $p 
\in P$, there is a QFA with $4 \frac {\log {2p}}{\epsilon}$ 
states recognizing $L_p$ with probability at least $1-\epsilon$. 
The proposed QFA construction depends on $d = 2\frac{\log 2p}{\epsilon}$ parameters $k_1, \dots, k_d$ 
and accepts input word $a^j \notin L_p$ with probability 
\[ \frac{1}{d^2} \left(\sum_{i=1}^{d} \cos \frac{2\pi k_i j}{p}\right)^2 .\]
It is possible to choose $k_1, \dots, k_d$ values to ensure  
\[ \frac{1}{d^2} \left(\sum_{i=1}^{d} \cos \frac{2\pi k_i j}{p}\right)^2 < \epsilon\] 
or, equivalently,
\begin{equation}
\label{eq:req} \left| \sum_{i=1}^{d} \cos \frac{2\pi k_i j}{p} \right| < 
\sqrt{\epsilon} d
\end{equation} for every $a^j \notin L_p$.

However, our proof is by a probabilistic argument and does not give an 
explicit sequence $k_1, \ldots, k_d$. We now present two constructions of
explicit sequences. The first construction works well in numerical experiments
and gives a QFA with $O(\log p)$ states in all the cases that we tested.
The second construction uses a slightly larger number of states but has
a rigorous proof of correctness.

\subsection{The first construction: cyclic sequences}
\label{sub:cyclic}

We conjecture

\begin{Hypothesis}
If $g$ is a primitive root modulo $p \in P$, then sequence $S_g = \{k_i 
\equiv g^i \bmod p\}_{i=1}^d$ for all $d$ and all $j: a^j\notin L_p$ 
satisfies (\ref{eq:req}).
\end{Hypothesis}

We will call $g$ a {\em sequence generator}. The corresponding sequence 
will be referred as cyclic sequence. We have checked all 
$p \in \{2, \ldots, 9973\}$, all generators $g$ and  
all sequence lengths $d < p$ (choosing a corresponding $\epsilon$ value) 
and haven't found any counterexample to our hypothesis.

We now describe numerical experiments comparing two strategies:
using a random sequence $k_1, \ldots, k_d$ and using a cyclic 
sequence.

We will use $S_{rand}$ to denote random sequence 
and $S_g$ to denote a cyclic sequence with generator $g$. 
We will also use $\epsilon_{rand}$ and $\epsilon_g$ to denote the maximum
probability with which a corresponding automata accepts input word $a^j 
\notin L_p$.

Table \ref{tab:sequence_examples} shows $\epsilon_{rand}$ 
and $\epsilon_g$ for different $p$ and $g$ values. 
$\epsilon_{rand}$ is calculated as an average over 
5000 randomly selected sequences.
$\epsilon_g$ is for one specific generator.
$\epsilon$ in the second column shows the theoretical 
upper bound given by Theorem \ref{T6}.

\begin{table}[h]
\centering
\begin{tabular}[b]{| c | c | c | c | c | c |}

\hline
$p$ & $\epsilon$ & $d$  & $g$ & $\epsilon_{rand}$ & $\epsilon_{g}$ \\

\hline
1523	& 0,1	& 161	& 948		& 0,03635	& 0,01517	\\
2689	& 0,1	& 172	& 656		& 0,03767	& 0,01950	\\
3671	& 0,1	& 179	& 2134	& 0,03803	& 0,02122	\\
4093	& 0,1	& 181	& 772		& 0,03822	& 0,01803	\\
5861	& 0,1	& 188	& 2190	& 0,03898	& 0,01825	\\
6247	& 0,1	& 189	& 406		& 0,03922	& 0,02006	\\
7481	& 0,1	& 193	& 6978	& 0,03932	& 0,01691	\\
8581	& 0,1	& 196	& 5567	& 0,03942	& 0,02057	\\
9883	& 0,1	& 198	& 1260	& 0,04011	& 0,01905	\\
\hline

\end{tabular}
\caption { $\epsilon_{rand}$ and $\epsilon_g$ for different $p$ and $g$}
\label{tab:sequence_examples}
\end{table}


In 99.98\% - 99.99\% of our experiments, random sequences 
achieved the bound of Theorem \ref{T6}. 
Surprisingly, cyclic sequences substantially outperform random ones
in almost all the cases.

\begin{figure}[h]
\centering
\includegraphics{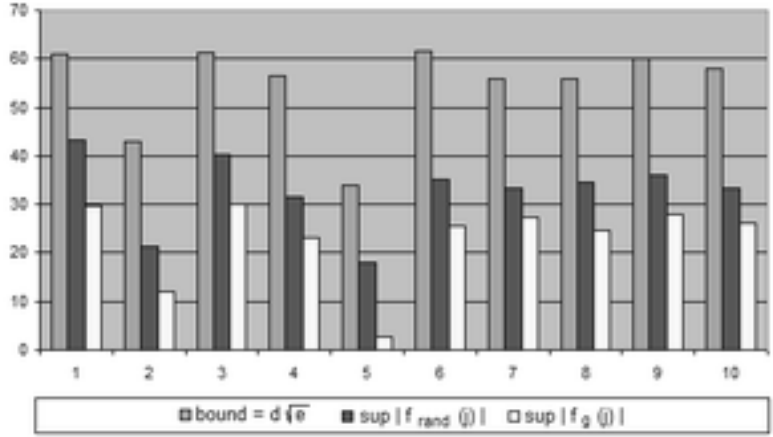}
\caption{$\sup | f_{g} (j) |$ and $\sup | f_{rand} (j) |$ for random $p, \epsilon$ and $g$}
\label{fig:F_g_wins}
\end{figure}

More precisely, for randomly 
selected $p \in P$, $\epsilon > 0$ and generator $g$, a cyclic 
sequence $S_{g}$ gives a better result than a random sequence $S_{rand}$
in 98.29\% of cases.
A few random instances are shown in Figure $\ref{fig:F_g_wins}$.
For each instance, we show the bound $d\sqrt{\epsilon}$ on (\ref{eq:req})
obtained by a probabilistic argument, the maximum of $f_{rand}(j)$
(which is defined as the value of (\ref{eq:req}) for the sequence 
$S_{rand}$) over all $j$, $a^j\notin L_p$ and the maximum of $f_g(j)$
(defined in a similar way using $S_g$ instead of $S_{rand}$).

\begin{figure}[h]
\centering
\includegraphics{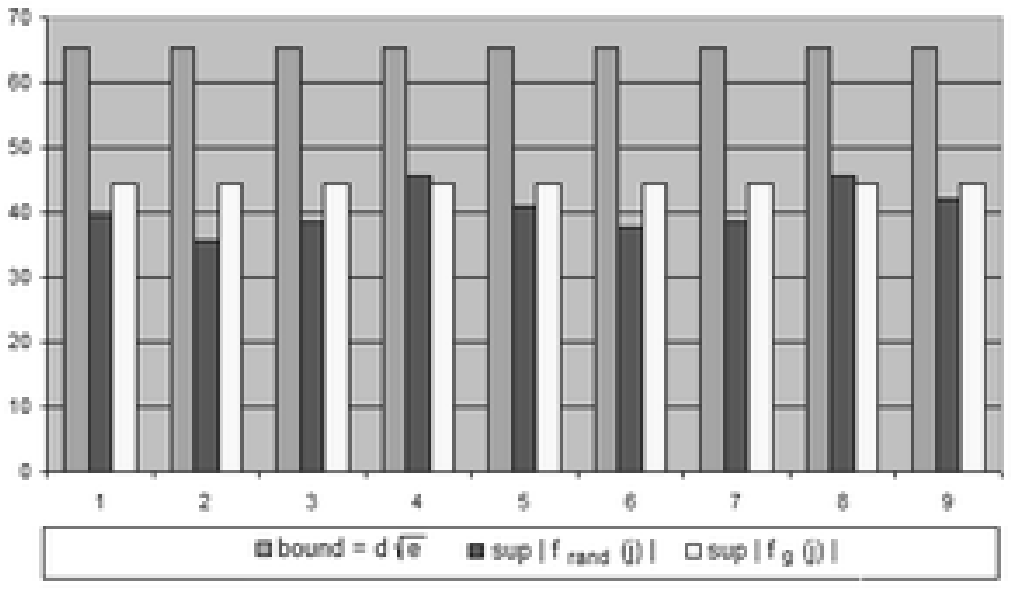}
\caption{$\sup | f_{g} (j) |$ and $\sup | f_{rand} (j) |$ for $p = 9059$, $\epsilon = 0.09$ and $g = 2689$}
\label{fig:F_rand_wins}
\end{figure}

In 1.81\% of cases, we got that $\sup | f_{g} 
(j) | > \sup | f_{rand} (j) |$, where $\sup | f_{rand} (j) |$ is 
calculated as an average over 5000 randomly selected sequences. 
Figure $\ref{fig:F_rand_wins}$ shows one of these cases: $p = 9059$, 
$\epsilon = 0.09$ and $g = 2689$, comparing the cyclic sequence with
9 different randomly chosen sequences. The cyclic sequence gives a 
slightly worse result than most of the random ones, but still beats
the probabilistic bound on (\ref{eq:req}) by a substantial amount.

\subsubsection{Comparing different generators}

Every $p \in P$ might have multiple generators. Table \ref{tab:different_generators} shows $\epsilon_{g}$ values for $p = 9059$ and $\epsilon = 0.1$ (sequence length $d = 197$, $\sqrt {\epsilon} d = 62.0101221453601$).

\begin{table}[h]
\centering
\begin{tabular}[b]{| c | c | c | c | c | c |}
\hline
$g$ & $\epsilon_{g}$ & $g$ & $\epsilon_{g}$ & $g$ & $\epsilon_{g}$\\

\hline
102 & 0,02533 & 1545 & 0,01858 & 9023 & 0,01807	\\
103 & 0,03758 & 1546 & 0,02235 & 9033 & 0,01413	\\
105 & 0,01999 & 1549 & 0,02896 & 9034 & 0,01485	\\
106 & 0,02852 & 1552 & 0,02873 & 9036 & 0,02509	\\
110 & 0,01685 & 1553 & 0,02624 & 9039 & 0,02311	\\
\hline

\end{tabular}
\caption { $\epsilon_g$ values for different generators. $p$ = 9059}
\label{tab:different_generators}
\end{table}

Different generators have different $\epsilon_{g}$ values. We will use $g_{min}$ to refer a minimal generator, i.e. one having a minimal $\epsilon_{g}$. Table $\ref{tab:min_generators}$ shows minimal generators for $p$ values from table $\ref{tab:sequence_examples}$.

\begin{table}[h]
\centering
\begin{tabular}[b]{| c | c | c | c | c | c | c |}

\hline
$p$ & $\epsilon$ & $d$ & $g$ & $\epsilon_{g}$ & $g_{min}$ & $\epsilon_{g_{min}}$ \\

\hline
1523	& 0,1	& 161	& 948		& 0,01517	& 624		& 0,00919	\\
2689	& 0,1	& 172	& 656		& 0,01950	& 1088	& 0,01060	\\
3671	& 0,1	& 179	& 2134	& 0,02122	& 1243	& 0,01121	\\
4093	& 0,1	& 181	& 772		& 0,01803	& 1063	& 0,01154	\\
5861	& 0,1	& 188	& 2190	& 0,01825	& 5732	& 0,01133	\\
6247	& 0,1	& 189	& 406		& 0,02006	& 97		& 0,01182	\\
7481	& 0,1	& 193	& 6978	& 0,01691	& 2865	& 0,01205	\\
8581	& 0,1	& 196	& 5567	& 0,02057	& 4362	& 0,01335	\\
9883	& 0,1	& 198	& 1260	& 0,01905	& 5675	& 0,01319	\\
\hline

\end{tabular}
\caption {Minimal generators for different $p$}
\label{tab:min_generators}
\end{table}

We see that, typically, the minimal generators give a QFA with 
substantially smaller probability of error. It remains open whether 
one could find a minimal generator without an exhaustive search of
all generators.

\subsection{The second construction: AIKPS sequences}

Fix $\epsilon>0$.
Let
\[ P=\{r | \mbox{$r$ is prime,~} (\log p)^{1+\epsilon}/2< r \leq (\log p)^{1+\epsilon} \},\]
\[ S=\{1, 2, \ldots, (\log p)^{1+2\epsilon}\} ,\]
\[ T=\{ s\cdot r^{-1} | r\in R, s\in S\} ,\]
with $r^{-1}$ being the inverse modulo $p$.
Ajtai et al. \cite{AIKPS} have shown

\begin{theorem}
\label{thm:ajtai}
\cite{AIKPS}
For all $k \in \{1, \ldots, p-1\}$,
\[ | \sum_{t\in T} e^{2 t k \pi i/p} | \leq (\log p)^{-\epsilon} |T| .\]
\end{theorem}

Razborov et al. \cite{RSW} have shown that powers $e^{2 t k \pi i/p}$ satisfy even 
stronger uniformity conditions. We, however, only need Theorem \ref{thm:ajtai}.

By taking the real part of the left hand side, we get
\[ \left| \sum_{t\in T} \cos\left( \frac{2 t k \pi i}{p} \right) \right| \leq (\log p)^{-\epsilon} |T| .\]
Thus, taking our construction of QFAs and using elements of $T$ as $k_1, \ldots, k_d$
gives an explicit construction of a QFA for our language with
$O(\log^{2+3\epsilon})$ states.

For our first, cyclic construction, the best provable
result is by applying a bound on exponential sums by Bourgain \cite{Bourgain}. 
That gives a QFA with $O(p^{c/\log \log p})$ states which is weaker than
both the numerical results and the rigorous construction in
this section.

{\bf Acknowledgment.}
We thank Igor Shparlinski for pointing out \cite{AIKPS} and \cite{Bourgain} to us.

\end{document}